\documentclass[ aps,%
floatfix,%
final,%
notitlepage,%
oneside,%
onecolumn,%
nobibnotes,%
nofootinbib,%
superscriptaddress,%
showpacs,%
11pt
]%
{revtex4-1}

\usepackage{amsmath}
\usepackage{graphicx}

\begin{document}

\large
\title{Production of $J/\psi$-meson pairs and $4c$-tetraquark at LHC}
\author{\firstname{A.~V.}~\surname{Berezhnoy}}
\email{Alexander.Berezhnoy@cern.ch}
\affiliation{SINP of Moscow State University, Moscow, Russia}
\author{\firstname{A.~K.}~\surname{Likhoded}}
\email{Anatolii.Likhoded@ihep.ru}
\affiliation{Institute for High Energy Physics, Protvino, Russia}
\author{\firstname{A.~V.}~\surname{Luchinsky}}
\email{Alexey.Luchinsky@ihep.ru}
\affiliation{Institute for High Energy Physics, Protvino, Russia}
\author{\firstname{A.~A.}~\surname{Novoselov}}
\email{Alexey.Novoselov@cern.ch}
\affiliation{Institute for High Energy Physics, Protvino, Russia}
\affiliation{Moscow Institute of Physics and Technology, Dolgoprudny, Russia}

\begin{abstract}
\normalsize
Theoretical predictions for $pp\to2J/\psi+X$ cross section at $\sqrt{s}=7$
TeV with different kinematical restrictions are presented. Results are 
compared with the the first LHCb data available. Special
attention is payed to possible signal from novel particles ---
tetraquarks build from two valence $c$-quarks and two valence $\bar{c}$-quarks.
According to our estimates it is quite possible to observe at least one of these states
(tensor tetraquark) experimentally under LHCb conditions.
\end{abstract}


\maketitle

\section{Introduction}

Simultaneous production of two $J/\psi$-mesons was first observed in 1982 by the NA3 collaboration
in multi-muon events in pion-platinum interactions at 150 and 280 GeV \cite{Badier:1982ae} and
later at 400 GeV in proton-platinum collisions \cite{Badier:1985ri}. Cross-section of this reaction
is suppressed compared to the single $J/\psi$ production by more than three orders of magnitude,
because of both higher order of perturbative QCD $O\left(\alpha_{s}^{4}\right)$ and kinematical reasons
due to production of bigger mass. At NA3 energies main contribution to the cross section arose from
quark-antiquark annihilation channel \cite{Kartvelishvili:1984ur}. On the contrary, it is the
$gg\to J/\psi J/\psi$ process studied in \cite{Humpert:1983yj} which dominates at the LHC.
Both contributions lead to the cross-section decreasing as the power of the invariant mass
of the $J/\psi$ pair ($\hat{\sigma}\sim1/\hat{s}^{3}$ for the $q\bar{q}$-annihilation
and $\hat{\sigma}\sim1/\hat{s}^{2}$ for the gluonic production). It is worth mentioning that such decrease
is absent in higher orders of perturbation theory and the cross-section of the $gg\to2J/\psi+X$
process becomes constant at large invariant masses \cite{Kiselev:1988ww}.
In current work we will be mainly interested in the low invariant mass region as it is most accessible
in the LHCb experiment, which specializes in heavy quarks studies.

In the leading
order in $\alpha_s$ quarkonia pair production process obeys selection rules similar to those
in quarkonia decays. Two initial gluons in a color-singlet state are C-even. That is why production
of $J/\psi$-, $\eta_c$- or $\chi_c$-meson pairs is allowed while combined production of 2
particles having different C-parity (such as $J/\psi+\eta_c$ and $J/\psi+\chi_c$) is prohibited.

First comparison of the NA3 data with QCD predictions has shown 2 times shortage of the prediction \cite{Kartvelishvili:1984ur}.
Taking additional contribution from the $P$-wave state $\chi_{c}$ or
$\psi'$-meson decay to $J/\psi$ into account could improve the agreement. Selection rules mentioned
above put restrictions on the feed down from the higher states. For instance feed down from the
$J/\psi+\chi_{c}$ production is expected to be less than from the $\chi_{c}+\chi_{c}$ one as production of the
former final state is prohibited in leading order. Meanwhile feed down from the
$J/\psi+\psi'$ channel is expected to be quite large. In principle, other mechanisms
of $J/\psi$-pair production are possible at the same order in $\alpha_{s}$. First of all it is contribution
of the color-octet states which plays significant role in the single $J/\psi$ production in the high-$p_T$ region.
Cross section of the color-octet states production falls more slowly as transverse momenta grows than
that of the color-singlet one. However at small transverse momenta and small invariant masses
of the $J/\psi$ pair smallness of the color-octet matrix element in the wave function of the $J/\psi$ meson compared
to the color-singlet one obviously leads to the negligible contribution. As shown in \cite{Qiao:2009kg,Ko:2010xy}
octet contribution becomes significant only at transverse momentum $p_{T}\gtrsim15$ GeV (which corresponds to
big invariant mass of the $J/\psi$ pair). Another possibility to enlarge theoretical prediction is connected with
the rearrangement of $c$-quarks in the final state, when instead of two colorless $c\bar{c}$-pairs two diquarks,
$cc$ and $\bar{c}\bar{c}$, are produced. After interaction in the final state they can form a pair of double-heavy
baryons above the threshold of their production or a pair of $J/\psi$-mesons below it.

The region of small invariant masses of the $J/\psi$-pair is most interesting due to the opportunity for two diquarks $[cc]_{\bar{3}_{c}}+\left[\bar{c}\bar{c}\right]_{3_{c}}$ to form a bound state --- tetraquark, which decays to a $J/\psi$-pair.
Attraction between $\bar{3}_{c}$- and $3_{c}$-states does not exclude such a possibility, especially since
similar exotic states like $Y(3940)$ resonance decaying to $J/\psi\omega$ \cite{Drenska:2010kg}
and $Y(4140)$ resonance decaying to $J/\psi\phi$ \cite{Wick:2010xv,Liu:2010hf} have recently been observed.

In addition, pair production of $J/\psi$-mesons has been earlier discussed as a possible way to observe C-even states
of bottmonia. Production of scalar and tensor $\chi_{b}$-mesons was considered in \cite{Braguta:2005gw,Kartvelishvili:1984en}, production
of an $\eta_{b}$-meson --- in \cite{Maltoni:2004hv}.

The second section of our article is devoted to the non-resonant production of $J/\psi$-meson pairs in the gluon-gluon interaction.
In the third section cross section of this process at LHC at $7$ TeV energy is calculated
taking different experimental restrictions into account. Special attention is payed to production in LHCb conditions
as there is first experimental data available \cite{exp}. Fourth section is devoted to calculation of the $4c$-tetraquark
mass and estimation of the cross section of its production at LHC.

\section{Double $J/\psi$-meson production in gluon-gluon interaction}

\begin{figure}
\begin{centering}
\includegraphics[width=16cm]{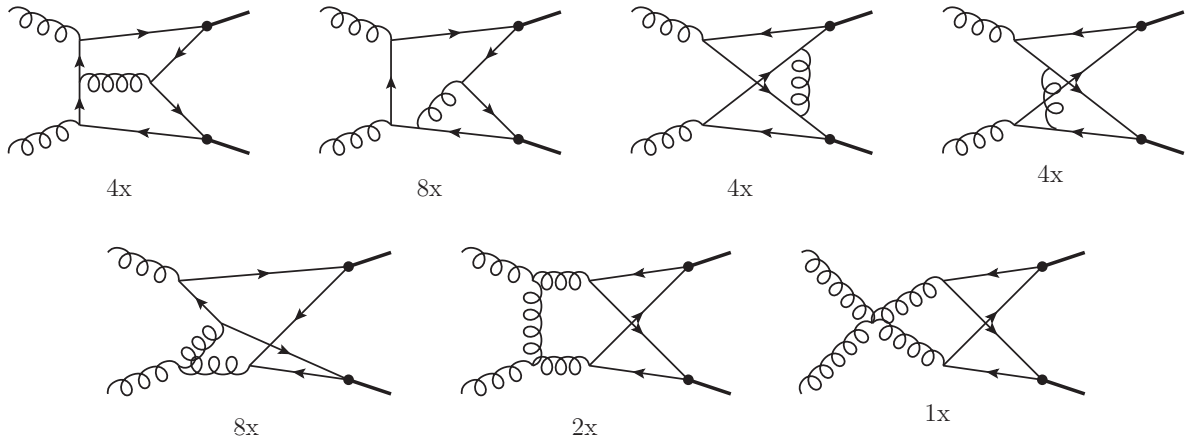}
\end{centering}
\caption{Feynman diagrams contributing $gg\to2J/\psi$ process. \label{fig:diags}}
\end{figure}

In the leading order of perturbative QCD there are 31 Feynman diagrams
describing color-singlet charmonium pairs production in gluonic reaction (see fig.\ref{fig:diags}).
We are not considering contribution of the quark-antiquark interaction, which is negligibly small at LHC energies.
Hadronization of a $c\bar{c}$-pair into a final $J/\psi$-meson is accounted for by the wave function of this
particle at origin:
\begin{eqnarray}
\left.\psi_{J/\psi}(r)\right|_{r=0} & = & 0.21\,\mathrm{GeV^{3/2}}.\label{eq:psiCC0}
\end{eqnarray}
It is the only nonperturbative parameter in the matrix element of $gg\to2J/\psi$ process. Its value is extracted from
the leptonic width of $J/\psi$-meson neglecting QCD corrections as we do not take these corrections into account in our matrix element.
Color-octet contributions in the kinematical region considered in our article can be neglected \cite{Qiao:2009kg,Ko:2010xy}.

In the following we will use two methods to calculate the cross section. First approach involves standard procedure
of analytical calculation of the amplitude, analytical squaring of this amplitude
(for example, using FeynCalc package~\cite{Mertig:1990an}) and subsequent integration over the phase space.
Second method which we use is based on numerical calculation of the amplitude at each point of the phase space,
followed by squaring (see papers \cite{Berezhnoy:1998aa,Berezhnoy:1996ks} for more details).
The latter method is associated with a smaller number of analytical calculations and makes it easier to consider
different quantum numbers of the final particles and their polarizations. Comparison of results
obtained by these two methods also help to check correctness of the calculations. Both methods give values
that agree with each other and coincide within errors of calculations with results presented in \cite{Humpert:1983yj,Li:2009ug}.

\begin{figure}
\begin{centering}
\includegraphics[width=10cm]{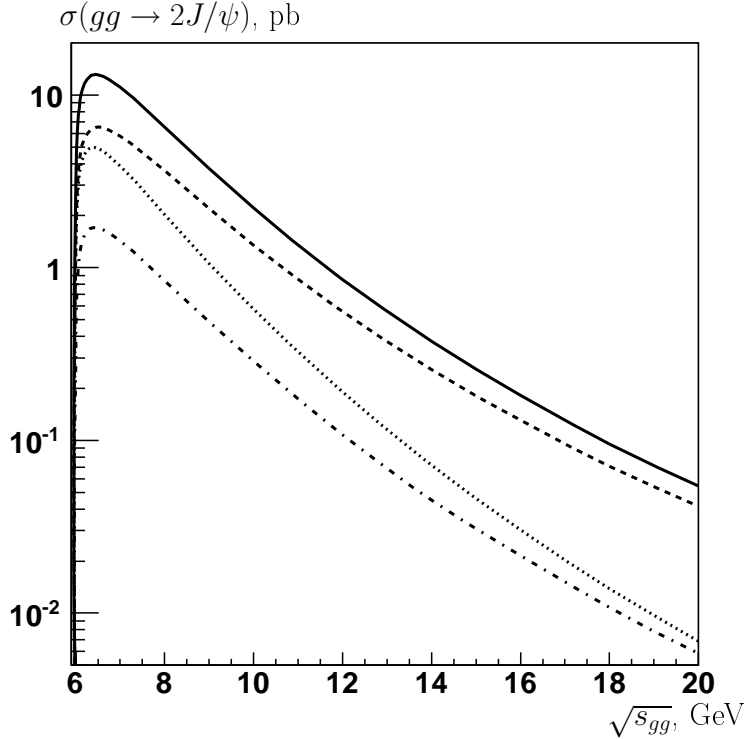}
\end{centering}
\caption{Cross section of double $J/\psi$-meson gluonic production as a
function of their invariant mass: solid curve --- total cross section;
dashed curve --- both mesons are transversely polarized;
dotted curve --- one meson is transversely polarized, another --- longitudinally;
dash-dotted curve --- both mesons are longitudinally polarized.
\label{fig:ggJJee}}
\end{figure}

\begin{figure}
\begin{centering}
\includegraphics[width=10cm]{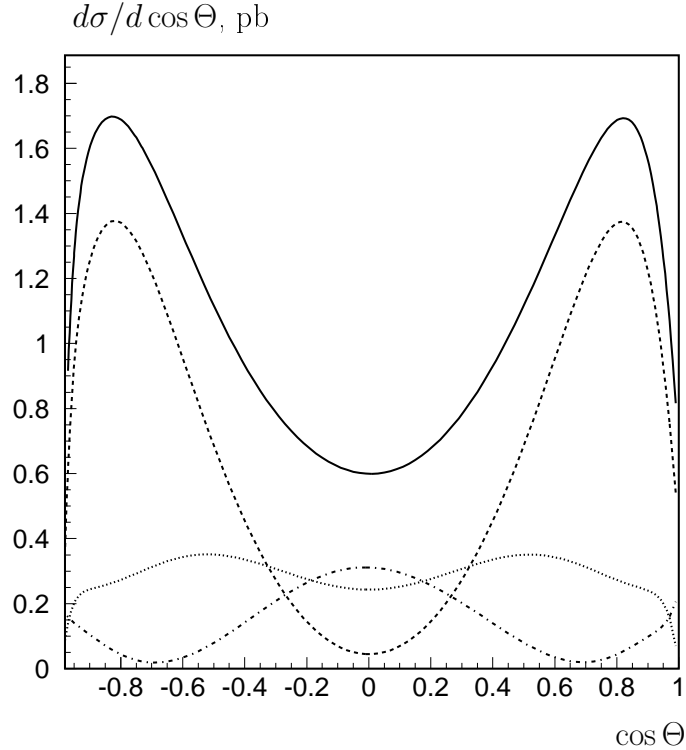}
\end{centering}
\caption{Angular distribution of different polarization states in $gg\to2J/\psi$
reaction at $\sqrt{\hat{s}}=10$ GeV: solid curve --- total cross section;
dashed curve --- both mesons are transversely polarized;
dotted curve --- one meson is transversely polarized, another --- longitudinally;
dash-dotted curve --- both mesons are longitudinally polarized.
\label{fig:angular}}
\end{figure}

In fig.\ref{fig:ggJJee} the cross section of double $J/\psi$-meson
production in gluonic interaction versus invariant mass of the pair is shown
for different combinations of polarizations. Table \ref{tab:polar}
shows cross section of the $J/\psi$ pair production for various polarizations
of each of $J/\psi$-particles. It can be seen that near
the threshold cross sections of transversely and longitudinally polarized
$J/\psi$-meson production are comparable, while at higher $p_{T}$
(and hence large invariant mass of the gluon pair), $J/\psi$-mesons
are mainly transversely polarized (see also \cite{Qiao:2009kg,Ko:2010xy}).
As an example, in fig.\ref{fig:angular} we show angular distribution
of polarized $J/\psi$-mesons, produced in $gg\to2J/\psi$ reaction
at $\sqrt{\hat{s}}=10$~GeV.
One can easily see, that angular distributions for different combinations
of polarizations differ drastically.

\begin{table}
\caption{Cross section of $gg\to2J/\psi$ processes for different combinations
of final particle's polarizations\label{tab:polar}}
\begin{centering}
\begin{tabular}{|c|c|c|c|}
\hline
$\sqrt{\hat{s}}$, GeV & $\hat{\sigma}(LL)$, pb & $\hat{\sigma}(LT)$, pb & $\hat{\sigma}(TT)$, pb\tabularnewline
\hline
\hline
6.2 & 1.59  & 4.66 & 5.75\tabularnewline
\hline
6.5 & 1.75 & 5.04 & 6.69\tabularnewline
\hline
7.0 & 1.47 & 3.99 & 5.96\tabularnewline
\hline
8.0 & 0.86 & 2.03 & 3.75\tabularnewline
\hline
9.0 & 0.50 & 1.09 & 2.26\tabularnewline
\hline
10.0 & 0.30 & 0.59 & 1.39\tabularnewline
\hline
12.0 & 0.11 & 0.196 & 0.57\tabularnewline
\hline
14.0 & 0.046 & 0.074 & 0.26\tabularnewline
\hline
16.0 & 0.022 & 0.031 & 0.136\tabularnewline
\hline
18.0 & 0.011 & 0.014 & 0.071\tabularnewline
\hline
20.0 & 0.006 & 0.0071 & 0.043\tabularnewline
\hline
\end{tabular}
\par\end{centering}
\end{table}

Partonic cross-section summed over polarizations as a function of $\hat{s}$
is equal
\begin{eqnarray}
\hat{\sigma}(\hat{s})&=&\frac{512 \pi ^3 \psi(0)^4 \alpha_s^4}{1215 m^4 \hat{s}^7} \times
\label{eq:partCS}
\\
\nonumber
&\times& \left( \hat{s} (5 \hat{s}^4 + 874 m^2 \hat{s}^3+ 13368 m^4 \hat{s}^2+36594 m^6 \hat{s}+ 90 m^8)
\sqrt{1-\frac{4 m^2}{\hat{s}}} + \right.
\\
\nonumber
&+& \left.
30 m^4 (6 m^6+70 m^4 \hat{s}-1025 m^2 \hat{s}^2-134 \hat{s}^3)\log \frac{1+\sqrt{1-\frac{4 m^2}{\hat{s}}}}{1-\sqrt{1-\frac{4 m^2}{\hat{s}}}}
\right)
\end{eqnarray}
Near the threshold there is a typical square-root dependence ($\hat{\sigma} \sim \sqrt{\hat{s}-4m_{J/\psi}^2}$),
which corresponds to the production of a pair in $S$-wave state. At large invariant masses expression (\ref{eq:partCS})
tends to $\sim \hat{s}^{-2}$ asymptotics.

As noted in the introduction $J/\psi$- and $\psi'$-meson or pair of $\psi'$-mesons can be produced
by the same mechanism in gluon-gluon interactions. According to dimensional considerations,
ratios of yields of different meson pairs are determined by the ratios
of their wave functions at origin and masses:
\begin{equation}
\sigma(2J/\psi) : \sigma(J/\psi+\psi') : \sigma(2\psi') \sim \frac{\psi_{J/\psi}^4(0)}{m_{J/\psi}^8} :
\frac {2 \psi_{J/\psi}^2(0)\psi_{\psi'}^2(0)}{((m_{J/\psi}+m_{\psi'})/2)^8} :
\frac{\psi_{\psi'}^4(0)}{m_{\psi'}^8}.
\end{equation}
The yield of non-identical particles is factor $2$ enhanced compared to
the yield of identical ones. We estimate contribution of the $gg \to J/\psi \psi'$ process by substituting
$(m(J/\psi)+m(\psi'))/2$ for $m(J/\psi)$ in the matrix element and using the
appropriate $\psi(0)$ for the second quarkonium. For $\psi'$ this value
determined from the leptonic width is equal
\begin{equation}
\left.\psi_{\psi'}(r)\right|_{r=0} = 0.16\,\mathrm{GeV^{3/2}}.
\label{eq:psi2Sat0}
\end{equation}

Taking numerical values into account one gets approximately
\begin{equation}
\sigma(2J/\psi) : \sigma(J/\psi+\psi') : \sigma(2\psi') \sim 1 : 1/2 : 1/12.
\end{equation}

Noticing that branching fraction of $\psi' \to J/\psi +X$ decay is about
56\% one concludes that feed down from this excited state is nearly 30\%.

Determination of feed down from $P$-wave states like $\chi_c$ requires
calculation of another matrix element with $2$ $P$-wave
particles in the final state. Meanwhile production of $\chi_c + J/\psi$ states
is suppressed by C-parity conservation and feed down from $\chi_c$ pairs
is suppressed by branching squared. Naive estimation of feed down from the latter
process leads to the value of about $6\%$ in low invariant mass region.

\begin{figure}
\begin{centering}
\includegraphics[width=10cm]{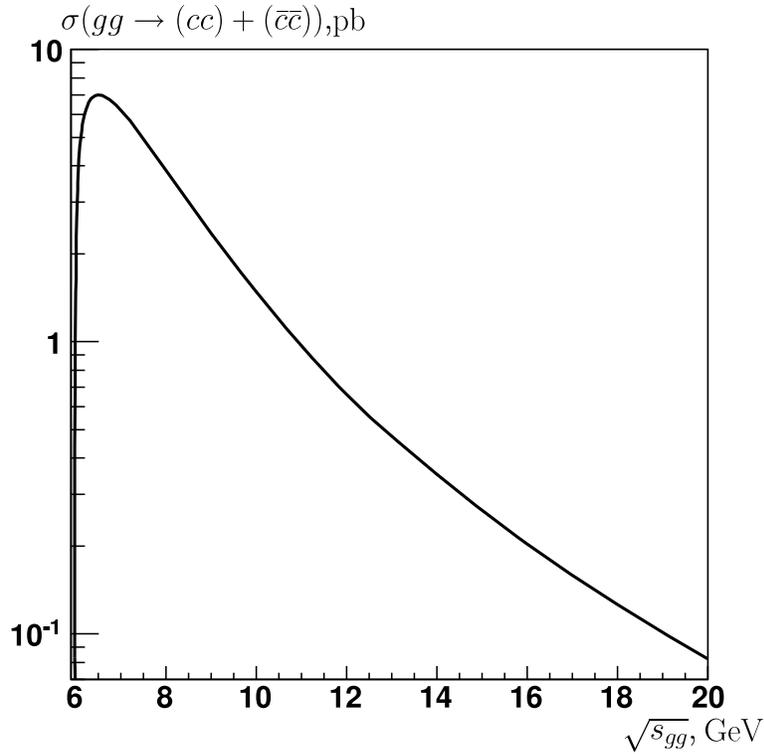}
\end{centering}
\caption{Cross section of diquark-antidiquark gluonic production as a
function of their invariant mass.
\label{fig:ggDD}}
\end{figure}

As noted above, near the threshold one of the sources of $J/\psi$-pair
is gluonic production of $cc$ and $\bar{c}\bar{c}$ diquarks.
This reaction is described by the diagrams similar to those shown in fig.\ref{fig:diags},
but with changed color factors and value of wave function at origin.
The last parameter can be calculated in the framework
of potential model. According to paper \cite{Kiselev:2002iy} its
numerical value is equal
\begin{equation}
\left.\psi_{[cc]}(r)\right|_{r=0} =  0.15\,\mbox{GeV}^{3/2}.
\label{eq:diqWFat0}
\end{equation}
In fig.\ref{fig:ggDD} the cross section of double diquark production
in gluonic interaction as a function of its invariant mass is shown.
Below the threshold of two $\Xi_{cc}$-baryons production at $m_{gg}\approx7$
GeV we expect the transition of this state into a pair of $J/\psi$-mesons
or a tetraquark, that can be observed as a peak in the $J/\psi J/\psi$
mass spectrum.

\section{Production at LHC}

To calculate
the cross section of charmonia pairs production in hadronic experiments
one needs to convolute partonic cross section presented in the
previous section with the distribution functions of partons in the
initial hadrons:
\begin{eqnarray}
d\sigma\left(pp\to2J/\psi+X\right)=\int dx_{1}dx_{2}f_{i}\left(x_{1}\right)f_{j}\left(x_{2}\right)d\sigma(ij\to2J/\psi),\label{eq:hard}
\end{eqnarray}
where $x_{1,2}$ are the momentum fractions carried by partons.
As already mentioned, gluon interaction gives main contribution
at the LHC energies, so it is sufficient to take only gluon fusion
into account.
Typical
values of $x$ are small, $x_{1,2}\sim10^{-3}$. According to our
calculations total cross section of the direct $J/\psi$ pairs production
in proton-proton interaction at $\sqrt{s}=7$ TeV is equal to
$$\sigma\left(pp\to2J/\psi+X\right) = 18\,\mbox{nb},$$
This value agrees well with previous results (\cite{Qiao:2009kg},
for example). Leading order expression for the running strong coupling constant
$\alpha_{s}(\hat{m}_T)$ was used to obtain this value.
It is clear that since the cross section
of the $gg\to2J/\psi$ process is proportional to $\alpha_{s}^{4}$,
final result depends strongly on the choice of this constant.
For gluon PDFs CTEQ5L~\cite{CTEQ} parametrization at $\hat{m}_T$ scale was used
as the most natural one for the leading order QCD process.

It worth mentioning that simple expression (\ref{eq:hard}) does not
account for the transverse momentum of gluons induced by the radiation
in the initial state (ISR). This effect obviously does not affect
value of the total cross section, but may be significant for the transverse
momentum distribution of the final particles as well as for the cross sections
in different detectors. In our work ISR
is taken into account by the Pythia 6.4 MC generator \cite{pythia}.
Standard tune of the version 6.4.25 was used.

\begin{figure}
\begin{centering}
\includegraphics[width=10cm]{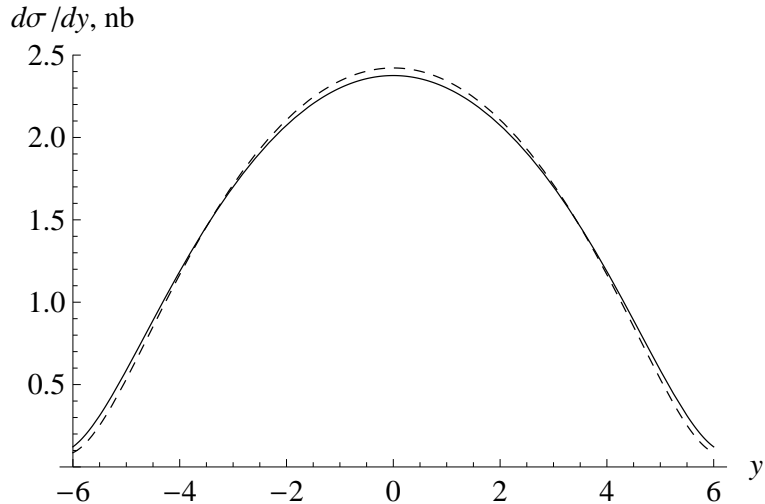}
\end{centering}
\caption{Rapidity distribution of the final charmonium in the reaction
$pp\to2J/\psi+X$.
Solid and dashed lines show the results obtained with the initial
state radiation accounted for and in the collinear approximation,
respectively.
\label{fig:ppY}}
\end{figure}

To obtain predictions for specific experiment it is necessary take
kinematical constraints imposed by the detector into account. For instance,
at the LHCb experiment main limitation is imposed on the rapidity
of the final charmonium: $2<y<4.5$, while there is virtually no
cutoff on the transverse momentum. Fig. \ref{fig:ppY} shows rapidity
distribution of the $J/\psi$-meson at the $\sqrt{s}=7$ TeV energy.
Solid and dashed lines correspond to the transverse motion of
partons accounted for and neglected, respectively. Cross section of
the $J/\psi$ pair production with the $2<y<4.5$ restriction imposed is equal
$$\sigma^{coll.}_{\mathrm{LHCb}}\left(pp\to2J/\psi+X\right) =  3.2\,\mbox{nb} $$
if the initial state radiation is neglected, and
$$\sigma_{\mathrm{LHCb}}\left(pp\to2J/\psi+X\right) = 3.1\, \mbox{nb} $$
if accounted for. One can see that refusal from the collinear approximation
affects cross section of the double $J/\psi$ production in LHCb very
slightly. This guaranties that uncertainty induced by the ISR effect
on the fraction of total cross section selected by the cutoff
is small compared to uncertainty in value of the total cross section itself.
The latter dues to $\alpha_s$, $\psi_{J/\psi}(0)$ and $m_c$ values and reaches 30\%.

\begin{figure}
\begin{centering}
\includegraphics[width=10cm]{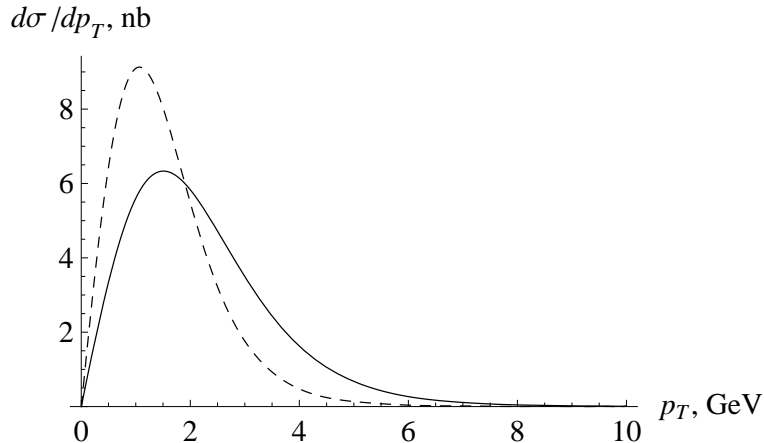}
\end{centering}
\caption{Transverse momentum distribution of the single $J/\psi$-meson obtained
with the initial state radiation accounted for (solid line) and in
the collinear approximation (dashed line).
\label{fig:ppPt}}
\end{figure}

The case is completely different for the detectors examining central
region and limiting minimal transverse momentum (such as ATLAS
and CMS). The restrictions on the transverse momentum and rapidity
of the final charmonia are:
$$ p_T > 5\,\mbox{GeV},\quad\left|y\right|<2.5. $$
Fig.~\ref{fig:ppPt} shows distributions over the transverse momentum
of one of the $J/\psi$-mesons obtained in the collinear approximation and
taking the ISR into account. It can be seen that accounting for the ISR
leads to a substantial broadening of the transverse
momentum distribution as compared to the collinear approximation.
According to our calculations cross section of the $J/\psi$ pair
production in the collinear approximation taking the ATLAS cutoffs into account equals
$$\sigma^{collinear}_{\mathrm{ATLAS}}\left(pp\to2J/\psi+X\right)  =  0.09\,\mbox{nb},$$
while accounting for the transverse motion of gluons lead (with the
same cutoffs) to the value
$$\sigma_{\mathrm{ATLAS}}\left(pp\to2J/\psi+X\right)  =  0.06\,\mbox{nb}.$$
Opposite to the situation in LHCb, this value is very model-dependent.
Moderate ISR decreases fraction of events selected by the cutoff,
as for two particles with transverse momenta slightly above the cutoff it
often moves one of them below the cutoff value. Intensive ISR
can increase this fraction as momenta of initial gluons more often become
compatible with the cutoff value and more particles exceed the cutoff. Meanwhile
behavior of the gluon's distribution over $p_T$ in the high $p_T$ region is
worst known. Values of the cross section under ATLAS conditions can differ
in several times depending on the tune of Pythia used.

\begin{figure}
\begin{centering}
\includegraphics[width=10cm]{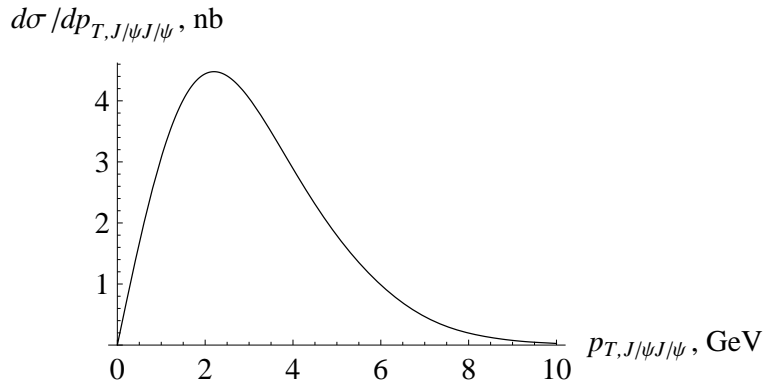}
\end{centering}
\caption{Distribution over the transverse momentum of the $J/\psi$ pair.
\label{fig:pTJJ}}
\end{figure}

To demonstrate importance of the initial state radiation we present
the $p_{T}$-distribution of the $J/\psi$-pair (fig.\ref{fig:pTJJ}).
In the collinear
approximation this distribution is obviously described by the $\delta$-function.

\begin{figure}
\begin{centering}
\includegraphics[width=10cm]{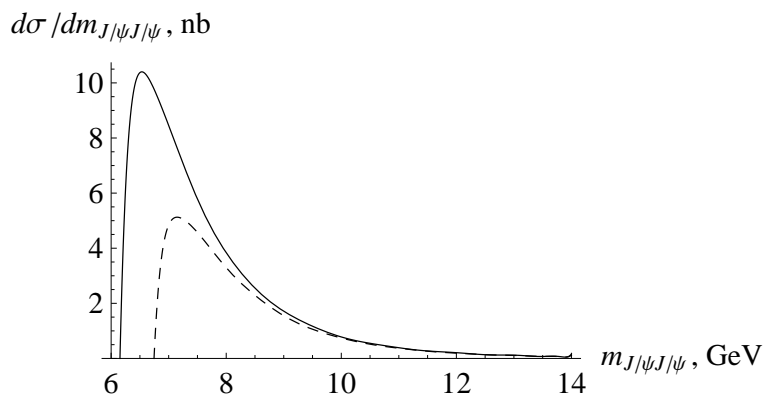}
\end{centering}
\caption{Distribution over the invariant mass of charmonium pairs in the
$pp\to J/\psi+X$ (solid curve) and $pp\to J/\psi+\psi'$ (dashed curve) reaction.
\label{fig:mJJ}}
\end{figure}

\begin{figure}
\begin{centering}
\includegraphics[width=10cm]{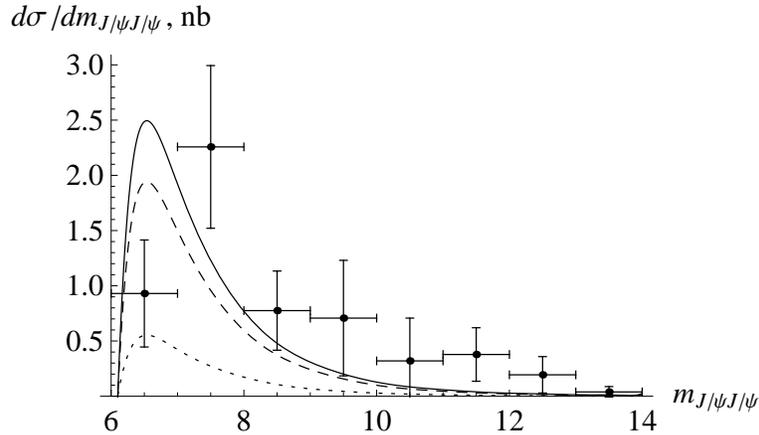}
\end{centering}
\caption{Distribution over the invariant mass of $J/\psi$ pairs produced
directly (dashed curve), with one $J/\psi$ originated from $\psi'$ decay (dotted curve)
and their total (solid curve). LHCb kinematical cutoff is applied.
Points with error bars correspond to the first LHCb data \cite{exp}.
\label{fig:mJJexp}}
\end{figure}

Solid curve in fig.~\ref{fig:mJJ} shows distribution over the invariant mass
of the $J/\psi$ pair. One can see that without cutoff in transverse momentum
main part of events is concentrated in the low invariant
mass region. Taking the transverse motion of the gluons into account
virtually does not affect shape of the invariant mass distribution,
as expected. Dashed curve in fig.~\ref{fig:mJJ} corresponds to the production
of the $J/\psi+\psi'$ final state. This distribution obviously starts at 
bigger invariant mass value than those of the $J/\psi$-pair. Meanwhile
invariant masses of $J/\psi$-pairs in which one of $J/\psi$-mesons originate
from the $\psi'$ decay ($\psi' \to J/\psi \pi \pi$ mostly) are distributed
similarly to direct $J/\psi$-pairs. Both distributions and their total
are presented in fig.~\ref{fig:mJJexp}. Kinematical cutoff corresponding
the LHCb detector ($2<y_{J/\psi}<4.5$) is applied to events saturating this distribution
to compare with the first LHCb data \cite{exp}.

Cutoff $2<y_{J/\psi}<4.5$ applied to the $pp \to J/\psi \psi' + X$, $\psi' \to J/\psi \pi \pi$
events simulated in Pythia generator selects approximately $16\%$ of the total feed down cross section,
which equals approximately $5.7$~nb. This corresponds to the feed down in LHCb equal to
$$\sigma_{\mathrm{LHCb}}^{\psi' \to J/\psi \pi \pi}\left(pp \to J/\psi \psi' + X\right) = 1.0\, \mbox{nb.} $$
Together with the direct cross section this leads to the value
$$\sigma_{\mathrm{LHCb}}^{total}\left(pp\to2J/\psi+X\right) = 4.1 \pm 1.2\, \mbox{nb.} $$
Experimental value reported in \cite{exp} is $5.6\pm1.1$nb, which is in a good agreement with the prediction.

\begin{figure}
\begin{centering}
\includegraphics[width=10cm]{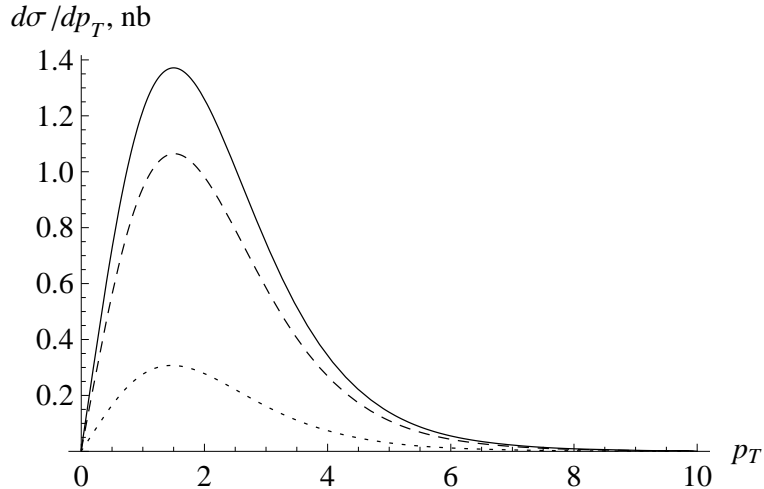}
\end{centering}
\caption{Distribution over the transverse momentum of a single $J/\psi$-meson from the pair produced
directly (dashed curve), with one $J/\psi$ originated from $\psi'$ decay (dotted curve)
and their total (solid curve). LHCb kinematical cutoff is applied.
\label{fig:pTexp}}
\end{figure}

\begin{figure}
\begin{centering}
\includegraphics[width=10cm]{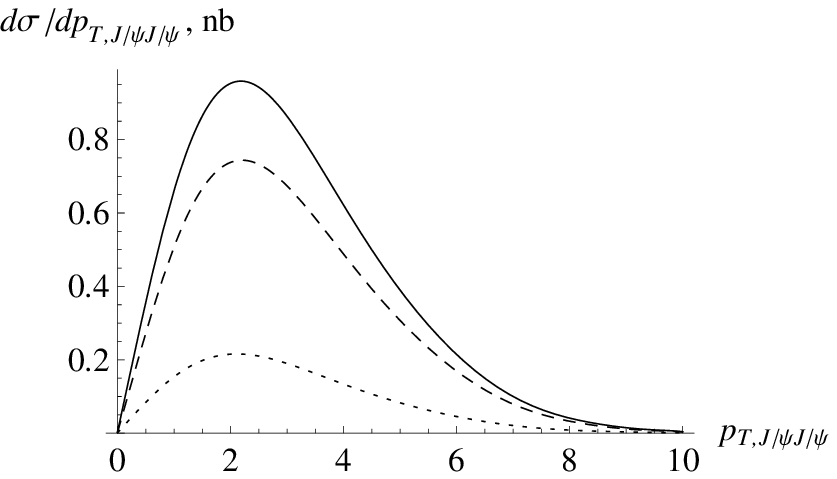}
\end{centering}
\caption{Distribution over the transverse momentum of a $J/\psi$-meson pair produced
directly (dashed curve), with one $J/\psi$ originated from $\psi'$ decay (dotted curve)
and their total (solid curve). LHCb kinematical cutoff is applied.
\label{fig:ptJJexp}}
\end{figure}

\begin{figure}
\begin{centering}
\includegraphics[width=10cm]{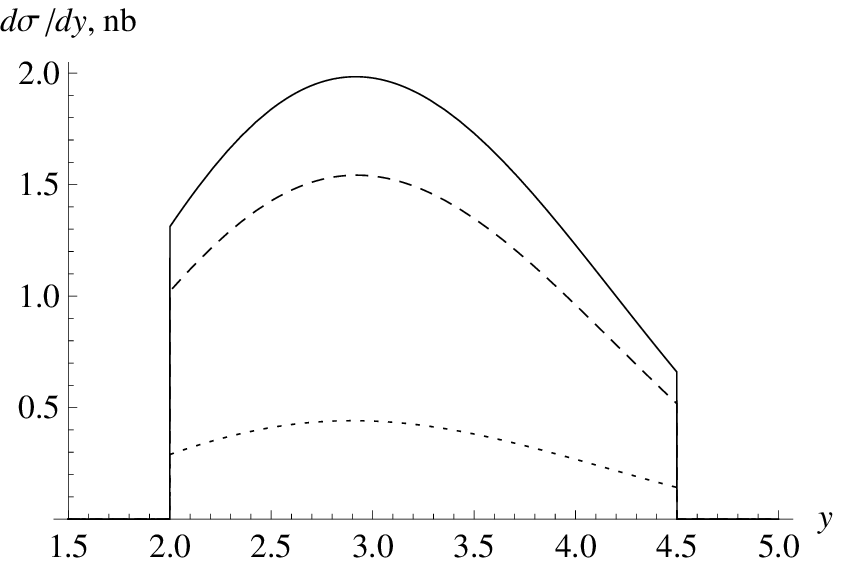}
\end{centering}
\caption{Distribution over the rapidity of a single $J/\psi$-meson of the pair produced
directly (dashed curve), with one $J/\psi$ originated from $\psi'$ decay (dotted curve)
and their total (solid curve). LHCb kinematical cutoff is applied.
\label{fig:Yexp}}
\end{figure}

We make predictions for different kinematical distributions which will be measured by LHCb later
at bigger statistics. Figures \ref{fig:pTexp} and \ref{fig:ptJJexp} show
distributions over transverse momentum of a single $J/\psi$ from a pair and over transverse momentum
of the whole pair respectively. Contributions from both direct production and
from $\psi'$ decays are shown. One sees that shape of both contributions are similar.
Fig. \ref{fig:Yexp} shows distribution over the rapidity of one of the $J/\psi$ mesons from a pair.
Smooth left border of this distribution dues to the fact that two $J/\psi$ mesons in a pair are generally close in rapidity.
So when one of them fails cutoff in rapidity, second one is rejected too, even if it is in the allowed range.

\section{Tetra-c-quark}

The most interesting phenomenon in the region of small invariant masses of $J/\psi$-pairs
is formation of a tetraquark built of 2 $c$-quarks and 2 $\bar{c}$-quarks and its decay
into a $J/\psi$-pair. Similar states, for example exotic mesons
$X(3872)$, $X(3940)$, $Y(3940)$ and $Y(4140)$, have already
been observed in $B$-meson decays and $J/\psi\phi$, $J/\psi\omega$ invariant mass
spectrums \cite{Wick:2010xv,Liu:2010hf}.
These states have small widths and can not be described in usual
quark-antiquark scheme. It turns out, however, that all these mesons
can be described in terms of tetraquarks $\left[cq\right]_{\bar 3}\left[\bar{c}\bar{q}\right]_3$,
where $q$ is $u$-, $d$- or $s$-quark \cite{Drenska:2010kg,Gershtein:2007vi}.
In the following we consider the $q=c$ case, that is a tetraquark $T_{4c}$ built
from four charm quarks --- $[cc][\bar{c}\bar{c}]$.

To estimate parameters of these states we use diquark model of tetraquark
that is we assume that it is built from two almost point-like diquarks
which interact with each other.
In this case Pauli exclusion principle
imposes strong restrictions on the quantum numbers of these diquarks.
Indeed, the angular momentum of quarks in the ground state of the
diquark should be equal to zero.
In order to attract to each other and build a tetraquark state both diquark and
antidiquark should be in the triplet color state. From the Pauli exclusion
principle it immediately follows that the total spin of both diquarks
must be equal to unity. Indeed coordinate component of wave function
of these diquarks is symmetric, color component --- antisymmetric, so
spin component of wave function has to be symmetric.
Mass of a diquark with such quantum numbers can be determined by solving
a non-relativistic Schrodinger equation \cite{Kiselev:2002iy,Gershtein:2000nx}.
Interaction potential is chosen proportional to the potential
between quark and antiquark in usual charmonium states but with an
additional factor $1/2$ which accounts for the difference in color
structures.
Using mass of the $c$-quark equal to $m_{c}=1.468$~GeV and taking
into account hyperfine splitting
$$
\Delta m  =  \frac{16\pi\alpha_s}{9m_c^2}(\mathbf{S_{c1}S_{c2}})\left|\psi_{[cc]}(0)\right|^2\sim 6.4 \,\mbox{MeV},
$$
where $\mathbf{S_{c1,2}}$ are quark spin operators and $\psi_{[cc]}(0)$ is the diquark
wave function at origin, which value was given in expression (\ref{eq:diqWFat0}),
the following values for diquark mass and mean radius can be obtained:
$$
m_{[cc]}  =  3.13\,\mbox{GeV},\qquad\left\langle r_{[cc]}\right\rangle =0.523\,\mathrm{\text{fm}}.
$$

The following consideration assumes that both diquarks are point-like.
In this case tetraquark mass can be obtained by solving Schrodinger
equation with the potential used in meson spectrum calculation.
The wave function of tetraquark at origin appears to be equal
\begin{equation}
\left.\psi_{T_{4c}}(r)\right|_{r=0} = 0.47\,\mathrm{GeV^{3/2}}.
\end{equation}
Without spin-spin interaction mass and mean radius of tetraquark
are equal
\begin{equation}
M_{T_{4c}} =  6.12\,\mbox{GeV},\qquad\left\langle r_{T_{4c}}\right\rangle =0.29\,\mbox{fm}.\label{eq:tetra}
\end{equation}

It is seen that the mean radius of tetraquark and its constituent
diquarks are comparable, and therefore the values obtained for the
masses should be regarded as a rough estimation.
Finite size of the diquarks can be accounted for with the form factor
$$
F\left(r^{2}\right)  =  \exp\left\{ -\frac{r^{2}}{\left\langle r_{[cc]}\right\rangle ^{2}}\right\}.
$$
Our calculations however show that this correction increases diquark
mass only slightly, approximately by $40$~MeV.
Due to the spin-spin interaction, described by the potential
$$
V_{SS}(r)=\frac{32\pi}{9}\frac{\alpha_{s}}{m_{[cc]}^{2}}(\mathbf{S_{1}S_{2}})\delta(\mathbf{r}).
$$
the state with mass (\ref{eq:tetra}) splits into scalar, axial and tensor
tetraquarks with masses
\begin{eqnarray*}
J=0:\qquad M_{T_{4c}(0^{++})} & = & 5.97\,\mbox{GeV},\\
J=1:\qquad M_{T_{4c}(1^{+-})} & = & 6.05\,\mbox{GeV},\\
J=2:\qquad M_{T_{4c}(2^{++})} & = & 6.22\,\mbox{GeV}.
\end{eqnarray*}
It is evident that only tensor meson from this list is above the threshold
of $J/\psi$-meson pair production.
Corrections caused by the finite size of diquarks can, in principle,
slightly increase masses and move the axial meson above this threshold,
but its production in two gluon channel is forbidden by the charge parity conservation.
As for the scalar tetraquark its decay into two real $J/\psi$ mesons is
kinematically forbidden. One of the final $J/\psi$-mesons, however,
could be a virtual particle decaying into $\mu^{+}\mu^{-}$-pair.
For this reason it is very interesting to search for the resonance
in $J/\psi\mu^{+}\mu^{-}$ channel.

Duality relations can be used to estimate cross section of tetraquark
production. Let us consider reaction $gg\to [cc]_{\bar 3} + [\bar c \bar c]_3$
which cross section as a function of invariant mass of the system is plotted
in fig.\ref{fig:ggDD}. Below the threshold of $2$ doubly-heavy baryons $\Xi_{cc}$
diquarks can form a tetraquark $T_{4c}$ consisting of $4$ $c$-quarks with
subsequent decay into a $J/\psi$ meson pair. So, the following relation
should hold:
\begin{multline}
S=\int\limits _{2M_{J/\psi}}^{2M_{\Xi_{cc}}}dm_{gg} \sigma (gg\to X_{4c} \to 2J/\psi) \\ =  K \cdot \int\limits _{2M_{J/\psi}}^{2M_{\Xi_{cc}}}dm_{gg}\hat{\sigma}(gg\to [cc]_{\bar 3} + [\bar c \bar c]_3 ) \approx K \cdot 6.4\,\mbox{pb}\, \cdot \mbox{GeV},
\end{multline}
where $K<1$.

\begin{figure}
\begin{centering}
\includegraphics[width=10cm]{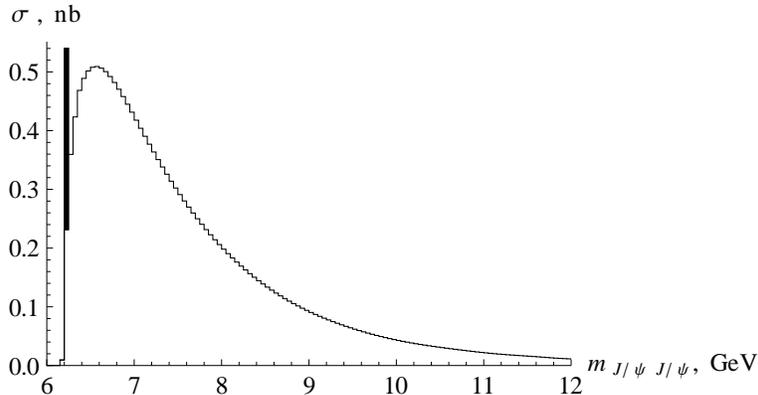}
\end{centering}
\caption{Tetraquark contribution to the distribution over the invariant
mass of the $J/\psi$-pair at 7TeV centre of mass energy.
\label{fig:tetra}}
\end{figure}

In the current work we use value $K=0.1$. Width and height of the peak caused by the
tetraquark contribution is determined by its own width $\Delta \sim 0.1~\mbox{MeV}$. 
However it is considerably
smaller than experimental resolution of the detector $\Delta_{\mathrm{exp.}} \sim 50~\mbox{MeV}$.
Histogram in Fig. \ref{fig:tetra} shows the distribution over the invariant mass
of the $J/\psi$-pair with expected contribution from
the tetraquark using experimental resolution for the bin width.

\section{Conclusion}

In our paper we use perturbative QCD to calculate the fourth order
contributions to the $J/\psi$ pairs production at LHC. According
to our estimates the ratio of the cross section of this reaction to
the cross section of single $J/\psi$-meson production in conditions
of the LHCb experiment ($2<y<4.5$) is about $4\times10^{-4}$.
It should be noted, that comparison of the first experimental
results at low energies with theoretical predictions shown excess
of experimental data by a factor $2-3$ \cite{Kartvelishvili:1984ur}.
Cross section predicted in our study is in a much better agreement 
with the recent LHCb result. Meanwhile all contributions accounted for 
in our work lead to kinematical peak corresponding to the first 
bin in experimental data presented. Possible contributions from 
the diquark-antidiquark production also lay in this invariant mass 
interval. If further experimental analysis eliminates lack of signal
in the first bin preserving shape of the distribution corresponding
increase of the cross section measured can amount to $50\%-100\%$.

It is well known, that production and subsequent radiative decays
of $P$-wave states $\chi_{c}$ can give additional contribution
to $J/\psi$-meson cross section. Our estimates in the region of low invariant
masses of the $J/\psi$ pairs do not support this view, though
it requires additional quantitative validation. Contribution form 
$\chi_{c}$ states can however become significant at large invariant 
masses and it is important to distinguish it from the color-octet one. The 
latter also is hardly noticeable in the kinematical region considered in
our article. Other possible source of enhancement 
is inclusive production of $J/\psi$-pairs accompanied by additional
gluons. Despite the fact that this process is described in the higher
orders of perturbation theory, the number of color degrees of freedom
in the final state increases greatly. It does not seem to be possible
to distinguish experimentally exclusive $J/\psi$-pair production and 
reactions with emission of additional gluons. One can only estimate 
possible enhancement caused by contributions from higher orders of QCD. 
Thus accurate measurement of the cross section of double $J/\psi$ 
production is needed.

Another interesting question considered is the possibility of observation of
a new exotic meson built from 2 $c$-quarks and 2 $\bar c$-quarks.
Our estimates show that if diquarks $\left[cc\right]$ and $\left[\bar{c}\bar{c}\right]$
are both in color-triplet states than in addition to attraction of quarks in diquark
mutual attraction of diquarks exist. Rough estimation of masses of resulting tetraquarks 
is presented in our paper and in is shown that at least one of these states (tensor tetraquark) can
be observed in $J/\psi J/\psi$ or $J/\psi\mu^{+}\mu^{-}$ modes.

Authors would like to thank V.V. Kiselev for the program for bound
state masses calculation and I. Belyaev for fruitful
discussions. The article was financially supported by Russian Foundation
for Basic Research (grant \#10-02-00061a). The work of A.V. Luchinsky
and A.A. Novoselov was also supported by non-commercial foundation
``Dynasty'' and the grant of the president of
Russian Federation (grant \#MK-406.2010.2).

\end{document}